\documentclass[preprint,12pt]{elsarticle}

\usepackage{graphicx}
\usepackage{amssymb}
\usepackage{lineno}
\usepackage{enumitem}
\usepackage{amsmath}
\usepackage[colorlinks]{hyperref}

\begin{document}

\begin{frontmatter}


\title{Rational hyperbolic discounting}



\author{Jos\'{e} Cl\'{a}udio do Nascimento}

\address{Universidade Federal do Cear\'{a}, Brazil}

\begin{abstract}
How much should you receive in a week to be indifferent to \$ 100 in six months? Note that the indifference requires a rule to ensure the similarity between early and late payments. Assuming that rational individuals have low accuracy, then the following rule is valid: if the amounts to be paid are much less than the personal wealth, then the $q$-exponential discounting guarantees indifference in several periods. Thus, the discounting can be interpolated between hyperbolic and exponential functions due to the low accuracy to distinguish time averages when the payments have low impact on personal wealth. Therefore, there are physical conditions that allow the  hyperbolic discounting regardless psycho-behavioral assumption.
 \end{abstract}


\end{frontmatter}

%\linenumbers

\newpage
\tableofcontents
\newpage
\section{Introduction}

\begin{figure}[!ht]
	\centering
	\includegraphics[scale=0.65]{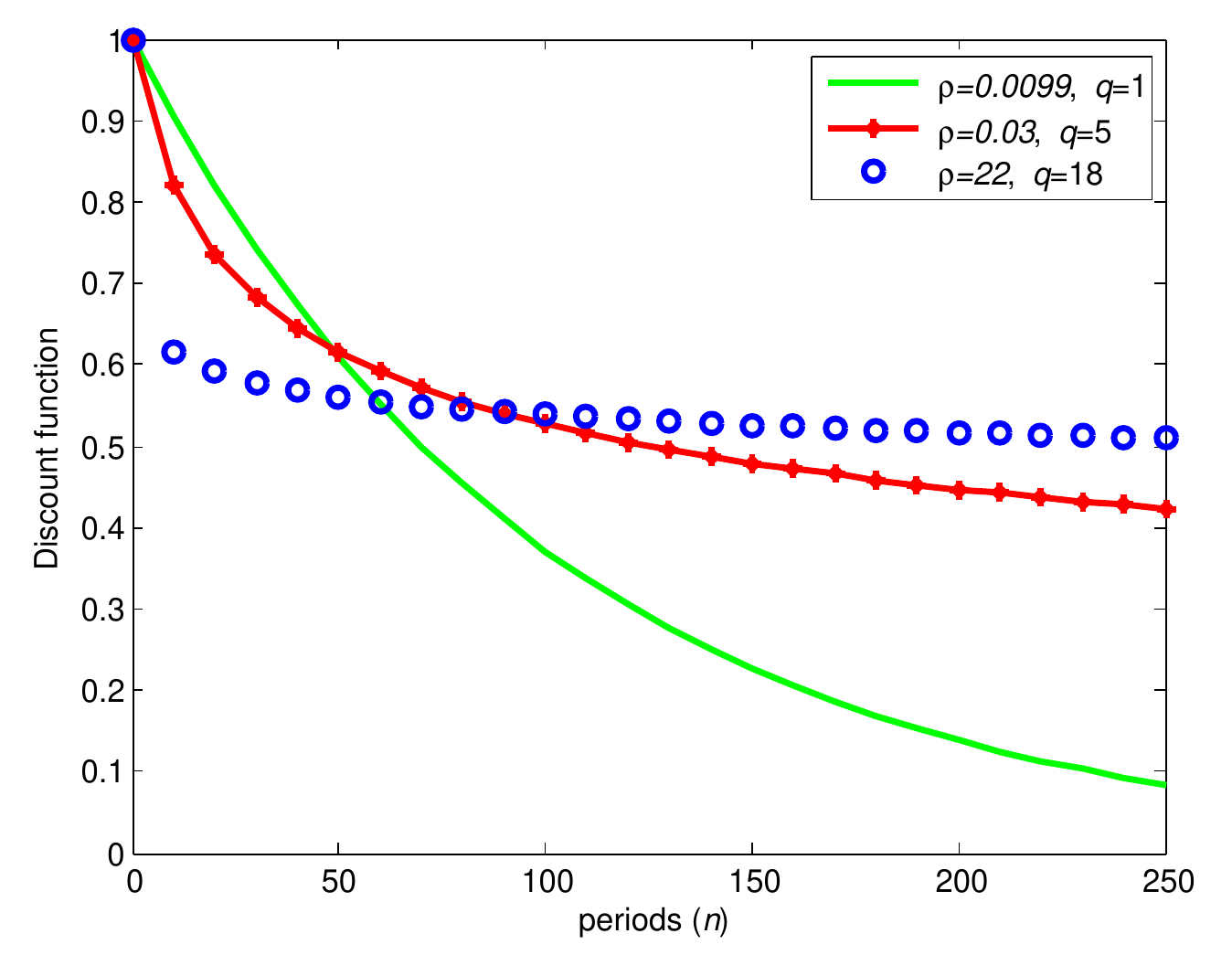}
		\caption{  Discount function $e_q^{-\rho n} $ versus  delayed periods. The green curve is the exponential discounting and the $\ast$-red curve is the hyperbolic discounting. The $\circ$-blue curve approximate the quasi-hyperbolic discounting (the fat tail is almost a straight line).}	
	\label{Fig:q_exponential}
\end{figure}

People are sensitive to  payment delays. To maintain indifference, a bigger amount should be paid in the future. To explain this problem, Paul Samuelson introduced the discounted utility model  \cite{samuelson1937note}, but experiments revealed behaviors that violate it \cite{frederick2002time}. For example, when mathematical functions are explicitly fitted to experimental data, then a hyperbolic shape fits the data better than the exponential form \cite{kirby1997bidding, kirby1995modeling, myerson1995discounting,rachlin1991subjective}. In general,  the $q$-exponential function
\begin{equation}
e_q^{-\rho n}\equiv \left[1-(1-q) \rho n \right]^{\frac{1}{1-q}}
\label{eq:q_exponential}
\end{equation}
 allows greater flexibility of fit for  hyperbolic, exponential, and quasi-hyperbolic discounting \cite{benhabib2004hyperbolic,takahashi2006time,cajueiro2006note,takahashi2007empirical} (see Figure \ref{Fig:q_exponential}). However, it is not clear whether the causes of the hyperbolic discounting are psychobehavioral or physical.

This paper evaluates the physical conditions to make an intertemporal choice, where  three realistic conditions are relevant: (1) impact of payments on personal wealth; (2) maximum uncertainty without testable information; (3) low accuracy on future amounts. Under these conditions, the time averages of early and late payments are calculated. Thus, it is possible to explain the preference reversal and discount problems.

The main result of this paper is in the evaluation of the discount conditions. More specifically, the discounting is strongly influenced by the impact on personal wealth and the low accuracy in the perception on future payments. In this physical context, if the payments are much less than the personal wealth, then the $q$-exponential discounting guarantees the indifference between early and late amounts.  Therefore,  the low accuracy to distinguish time averages allows different dynamics when the payments have low impact on personal wealth. This means that there is  physical causes to explain the hyperbolic discounting   regardless of psycho-behavioral assumptions.

This paper is organized as follows. In Section \ref{PTemp}, the time preference problem is presented and the necessary aspects to solve it are commented on. In Section \ref{SLT}, the time averages at the end of the first period are calculated evaluating similarities between  payment hypotheses. In Section \ref{DC}, a decision criterion to choose the highest time average is presented and the preference reversal problem is explained. In Section \ref{TCOIBEALP}, payments limited to a small fraction of the personal wealth generate low contrast and cause the hyperbolic discounting. In Section \ref{SOADM},  simulation of Thaler's experiment \cite{thaler1981some} is implemented with a personal wealth equal to $\$$ 100k. In Section \ref{DATWOTDM}, I discuss the pieces of evidence of my model in empirical observations. Finally, I provide some conclusions in Section \ref{Conclusion}.

\section{Time preference problem and its physical aspects}
\label{PTemp}

Time preference is the valuation placed on receiving a good on a short date compared with receiving other on a farther date.   Formally, given two hypotheses at present  about the receipt of payments (instant $t_0$),  then only one of them must be chosen by the decision maker, where 
\begin{enumerate}
\item the hypothesis $\Theta_m =$ ``I receive $m$'' represents the receipt of the amount $m$ at the end of a short period, $t_0+\delta t$, and
\item the hypothesis $\Theta_M =$ ``I receive $M$'' represents the receipt of the amount $M$ at the end of a long period, $t_0+\Delta t$. 
\end{enumerate} 
Therefore, there are $n$ short periods $\delta t$ until the given date to receive $ M $, $n=\Delta t/\delta t$.

The problem does not define all the conditions necessary to solve it. Therefore,   aspects of it must be examined to develop a solution approach.

\subsection{Impact on personal wealth}

 The  value of payment has an impact on personal wealth.  Particularly, if an individual has a  wealth $W_0$, then  a new wealth must be  calculated  when the proposition $\Theta_M$ is true. For this,  the impact factor $1+X_M$ is applied,  $W_1=W_0(1+X_M)$. Analogously,  $\Theta_m$ has the impact factor $1+X_m$. Thus, each growth rate is calculated as follows:  
\begin{equation}
X_m=m/W_0, \label{eq_impactM}  
\end{equation}
\begin{equation}
X_M=M/W_0. \label{eq_impactm}
\end{equation}
Therefore,  the growth rates $X_m$ and $X_M$ measure how intense is the impact on wealth.

\subsection{Repetition of exposures to risky payments}
\label{alertness}
  The time preference problem defines only one-time payments, but a repetition of exposures to risky payments should be considered.   The problem subjects a decision maker to a single experiment  as if  he  will never again take any equivalent risk \cite{taleb2018skin}. However, several experiments investigated discount rates of individuals in different contexts \cite{harrison2002estimating, warner2001personal, kirby1999heroin, coller1999eliciting, dreyfus1995rates, kirby1997bidding, kirby1995modeling, viscusi1989rates, moore1990models, moore1990discounting, gately1980individual, hausman1979individual}. This shows a diversity of situations where the time preference problem occurs, requiring ability to perceive new opportunities under uncertainty over time.
  Therefore,  a mode of repetition that transforms the time preference problem in a stochastic growth problem  needs to be assumed (see \cite{peters2016evaluating}).

\subsection{Foregone and chosen opportunity}

There is always a foregone opportunity and a chosen opportunity  after the decision. Each hypothesis represents an opportunity in a {\it possible world}\footnote{ The concept of a possible world is used to express modal statements, where a proposition is true at least one possible world \cite{kripke1959completeness, kripke1963semantical}. For example, if the decision maker chooses the possible world where he receives $M$, then the proposition $\Theta_m$ is denied in this world,  ``I don't receive $m$''. Finaly, the many-worlds idea is not strange in physics.  In quantum mechanics no superpositions of dead and alive cats are allowed in a single world \cite{dewitt2015many}}. Thus,  both hypotheses must be assessed as distinct processes in each world before making a decision. But after the decision, an opportunity is foregone and only the chosen opportunity will develop its sequence of irreversible facts.

\subsection{Payments probabilities and  maximum entropy}
\label{sub:payment_probabilities}

In this paper, a payment probability is the likelihood over a specified period to receive an amount. In other words, it is a cumulative probability over time. In general, there are three types of a priori cumulative density function to represent the waiting for payments:
\begin{enumerate}[label=(\alph*)]
\item degenerate distribution is useful to represent deterministic payments, where payment delays and advances do not occur; 
\item specific distribution for a given context where the information is testable;
\item uniform distribution can represent random payments within time ranges when there is not testable information. For example, we can write the cumulative probability $P_M(t)$ by
\begin{equation}
    P_M(t)=\left\{\begin{array}{ll}
0&\mbox{for}\quad t< t_0,\\
\frac{t-t_0}{\alpha\Delta t} &\mbox{for}\quad t_0\leq t \leq t_0+\alpha\Delta t\quad \mbox{and}\quad \alpha\geq 1\\
1 & \mbox{for}\quad  t > t_0+\alpha\Delta t
\end{array}\right.
\end{equation}
In this case, note that $P_M(t_0+\delta t)=\frac{1}{\alpha n}$.
\end{enumerate}

\begin{figure}[!ht]
	\centering
	\includegraphics[scale=0.60]{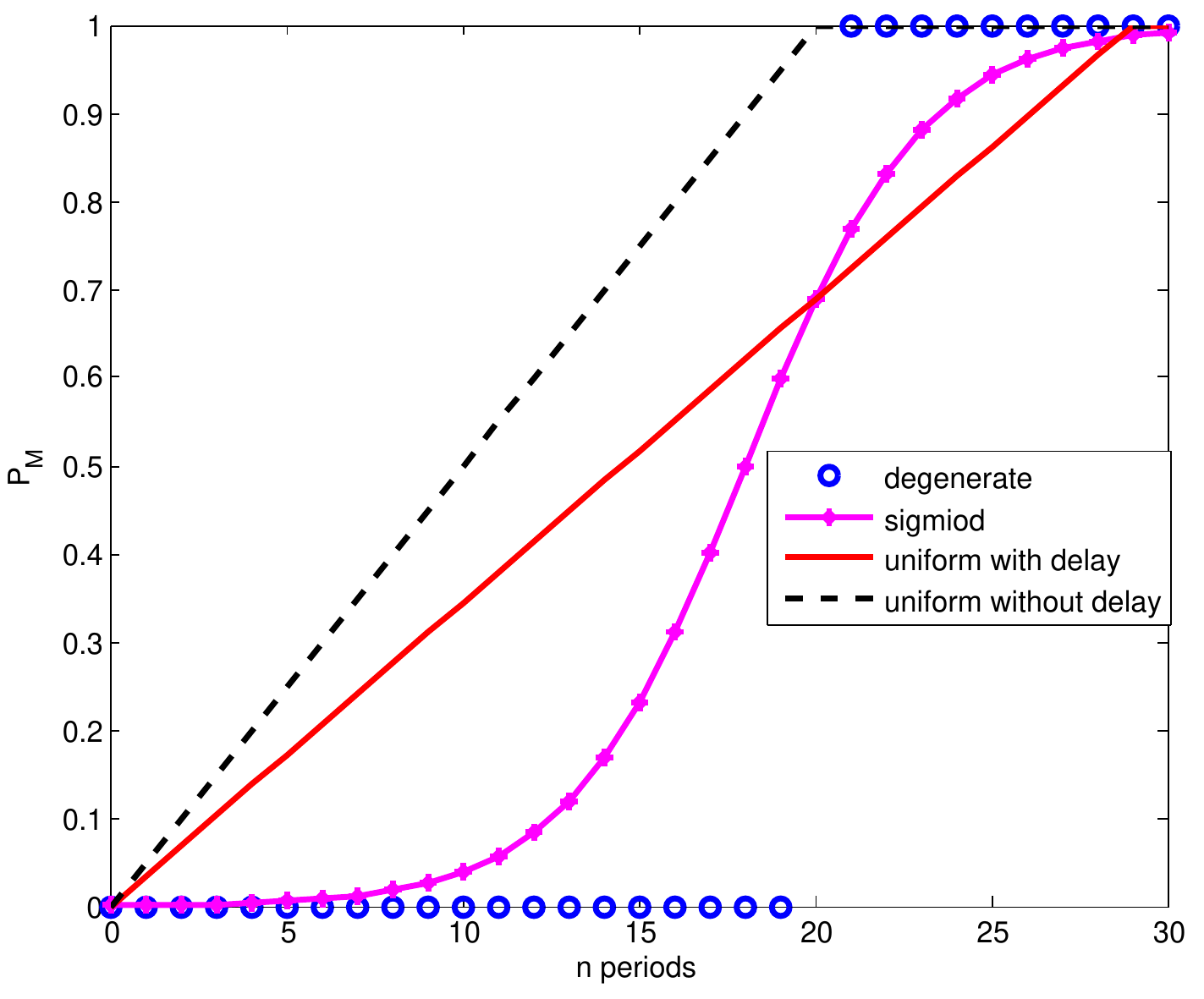}
		\caption{Cumulative density functions for payments at the end of 20 periods: 
		(a)  the degenerate distribution represents a deterministic payment ($\circ$-blue line); 
		(b) Many common probability distributions have the sigmoidal form ($\ast$-margent curve). An example is the normal distribution. However, sigmoidal form is not mandatory. In general situations, the payment probability distribution is unknown;  
		(c) The uniform distribution has maximum entropy when there is not testable information. The dashed-black line has cumulative probability equal to 1 at $n=20$ (uniform distribution without delay, $\alpha =1$). On the other hand, the red line has cumulative probability equal to 1 after 20 periods, therefore, there is a delay, $\alpha >1$. }	
	\label{Distributions}
\end{figure}

 Figure \ref{Distributions} shows some cumulative density functions. Initially, see the degenerate distribution ($\circ$-blue line). The probability is zero in any period before the payment date, but it is 1 at the payment moment and on the following periods. Thus, it communicates that the event has already occurred after the scheduled date. This type of distribution is applied only to deterministic payments, so it is outside the scope of this paper.

On the other hand, when there are risky payments, a probability distribution describes the context of uncertainty with a smoother curve than that observed in the degenerate distribution (see the $\ast$-margent sigmoid as a fictional example  in Figure \ref{Distributions}). However, obtaining testable information is not feasible in this case. After all, it is hard to discover the real probability distribution in several context of payments. Therefore, the entropy maximization must at least respect the unit measure (sum of all probabilities equal to 1). Under this constraint, uniform distribution maximizes the uncertainty \cite{jaynes1957information}.

 If the distribution is uniform, then the uncertainty depends on the time range. Therefore, if payments can be delayed beyond the deadlines, then the $P_M$ function increases its entropy reaching the value 1 later (compare $\alpha=1$ and $\alpha>1$, dashed black and red lines in Figure \ref{Distributions}).

 The cumulative probabilities $P_m$ and $P_M$ can describe  the payment probabilities over any periods. However, most of this paper  refers to payments at the end of the first period, $t_0 + \delta t$. So, for the sake of notational simplicity, $p_m$ and $p_M$ refer to the probability at this moment,
\[ p_m=P_m(t_0 + \delta t) \quad\mbox{and}\quad
p_M=P_M(t_0 + \delta t).\]
 
\subsection{Certain payments}

In usual situations, small payments are more likely than large payments. Sometimes the earlier payment is considered a certain event. In this paper, both payments will be considered uncertain. Even so, we can propose hypotheses with  modest payments in order to obtain  quasi-deterministic payments in the first period, $t_0\leq t \leq t_0 + \delta t$. These modest hypotheses must have smaller impact factors, $1+x_m<1+X_m$ and $1+x_M<1+X_M$. In these cases, we express them by\\ \\
$\theta_m =$ ``I receive an amount less than $m$'' and \\ 
$\theta_M =$ ``I receive an amount less than $M$''.\\ \\
Since the $ \theta_m $ and $ \theta_M $  can be true at all periods, they are useful for counting the frequency which $ \Theta_m $ and $ \Theta_M $ are true.

\subsection{Fuzziness}

Finally,  decision makers are not perfect measurement devices. They have low accuracy to distinguish situations with low contrast. Therefore, we must identify whether a problem produces fuzziness or not for a decision maker. So, a technique for verifying the distinguishability of the information is required.

\section{Evaluating payment hypotheses}
\label{SLT}
 A  hypothesis is a proposition (or a group of propositions) provisionally anticipated as an explanation of facts, behaviors, or natural phenomena that must be later verified by deduction or experience. Formally, they should always be spoken or written in the present tense because they are referring to the research being conducted. However, in  everyday language, payment hypotheses are guesses for decision making before the receipts are verified.

Although payment hypotheses are presented in the present tense, the sense of certainty we have about them becomes evident when they are expressed in the future tense. Given that a future sentence indicates an event occurs after the moment of speech, then it can emphasize whether a fact will come true or not. For instance, consider the following future statements for  $\Theta_M$ and $\theta_M$:
 \\
 \\
 $F\Theta_M$ = ``It will at some time be the case that I receive   $M$'';\\
 $G\theta_M$ = ``It will always be the case that I receive an amount less than $M$''.
\\
\\
 The statement  $G\theta_M$ expresses  certainty with the modifier ``always'', while $F\Theta_M$ expresses uncertainty about the instant that the action occurs through the modifier ``some time''. In temporal logic,  $F\Theta_M$ is equivalent to saying which there is an instant  $t$ in the future where $\Theta_M$ is true, i.e, $\exists \;t$ such that (now $ <t$) $\wedge \Theta_M(t)$, where $\Theta_M (t)$ means that $\Theta_M$ is true at instant $t$. 
Meanwhile,  $G\theta_M$ affirms  that $\theta_M$ is always true in the future,  $ \theta_M(t) \;\; \forall \: t>$ now, \cite{prior2003time}. 

\subsection{Repetition of payments}
 As discussed previously, risky payments will emerge in different  situations, requiring {\it alertness}\footnote{The term alertness is formally used to refer the ability of entrepreneurs to perceive new opportunities in the market \cite{kirzner2015competition}. Here, this term refers to the ability to perceive new payment opportunities over time.} to take risk several times in the future (see subsection \ref{alertness}). Thus, there is a perspective that a payment of $M$ will be possible indefinitely times on the timeline. Specifically, we must  write
\\
\\
$GF\Theta_M$ = ``It will always be the case that I will at some time receive  $M$'',
\\
\\
where $GF\Theta_M$ states that the proposition $\Theta_M$    will be true several times in the future, but there will always uncertainty about the logic value in any  instant.

   Thus, different payment hypotheses can describe different dynamics when expressed in the future tense. For example, the chances of $\theta_M$ being true are greater than the chances of $\Theta_M$. On the other hand, the payments proposed in $\Theta_M$ are higher than those proposed in $\theta_M$. So that, the statement $GF\Theta_M$  communicates that the decision maker will frequently receive $M$   (impact factor  $1+X_M$). Meanwhile, $G\theta_M$  communicates the existence of payments in all future periods, where the impact factor $1+x_M$ must be small enough for that to be true.
  
   One of the speculation problems is generating hypotheses with different payments, but with similar performances when repeated over time. So, different performances can be compared to make a decision.   To solve this problem, we can count the future periods stated in  $GF\Theta_M$ and $G\theta_M$. Thus, let us consider $\tau(u)$ as the total time when $GF\Theta_M$ affirms that $\Theta_M$  is true, and $u$ as the total time when $G\theta_M$ affirms that $\theta_M$ is true. Then, the same performance is achieved over time when
  \begin{equation}
      (1+x_M)^u =(1+X_M)^{\tau(u)}.
      \label{eq:similarity1}
  \end{equation}
  Note that multiplicative dynamics is assumed in the above equation. The similarity between these two dynamics is denoted in this paper by
\begin{equation}
G {\theta_M}  \sim GF\Theta_M.
\end{equation}  

\subsection{One-time payments and the time average}
  Temporal logic allows us to remove the repetition of exposure to risky payments. The axiomatic system of temporal logic proposes that  
 $G {\theta_M} \Rightarrow GN {\theta_M}$, where  $N\theta_M$ affirms that $\theta_M$ will be true in the next period. Therefore, the similarity  $GN {\theta_M}  \sim GF\Theta_M   $ can be written by
\begin{equation}
N{\theta_M} \sim    F\Theta_M.
   \label{eq_est_MeiosisM}
\end{equation}
  
The similarity above indicates that the performances of the hypotheses are the same, even when the payments occur only once at different instants.  To verify this fact, we should note that $u$ is also the number of periods under observation because $G\theta_M$ affirms that $\theta_M$  will always be true (see equation \ref{eq:similarity1}). Then, $GF\Theta_M$ communicates that the action in $\Theta_M$ will occur with a frequency given by
\begin{equation}
\lim\limits_{u\to \infty} \frac{\tau(u)}{u}=p_M.
\label{eq:p}
\end{equation}
According to equation \ref{eq:p}, we have a stationary probability $p_M$ when $u$ is large enough.   Therefore, the relation between time  averages,\begin{equation}
    1+x_M = (1+X_M)^{p_M},
     \label{eq_MeiosisM}
\end{equation}
indicates that the sentences  $N\theta_M$   and $F\Theta_M$ have similar goals in the future, although $\theta_M$  and  $\Theta_M$ have different payments.  Specifically, the statement
``It will be the case that I receive an amount less than $M$ in the next period'', which has an impact factor $1+x_M$, is similar to the statement, ``It will at some time be the case that I receive   $M$'',  which have a   time average $(1+X_M)^{p_M}$. Analogously,  if  
\begin{equation}
   (1+x_m)= \left(1+X_m\right)^{p_m},
   \label{eq_Meiosism}
\end{equation}
then we  have $N\theta_m \sim F\Theta_m$, where $p_m$ is the payment probability to receive  $m$ on the next period.  

\begin{figure}[!ht]
	\centering
	\includegraphics[scale=0.60]{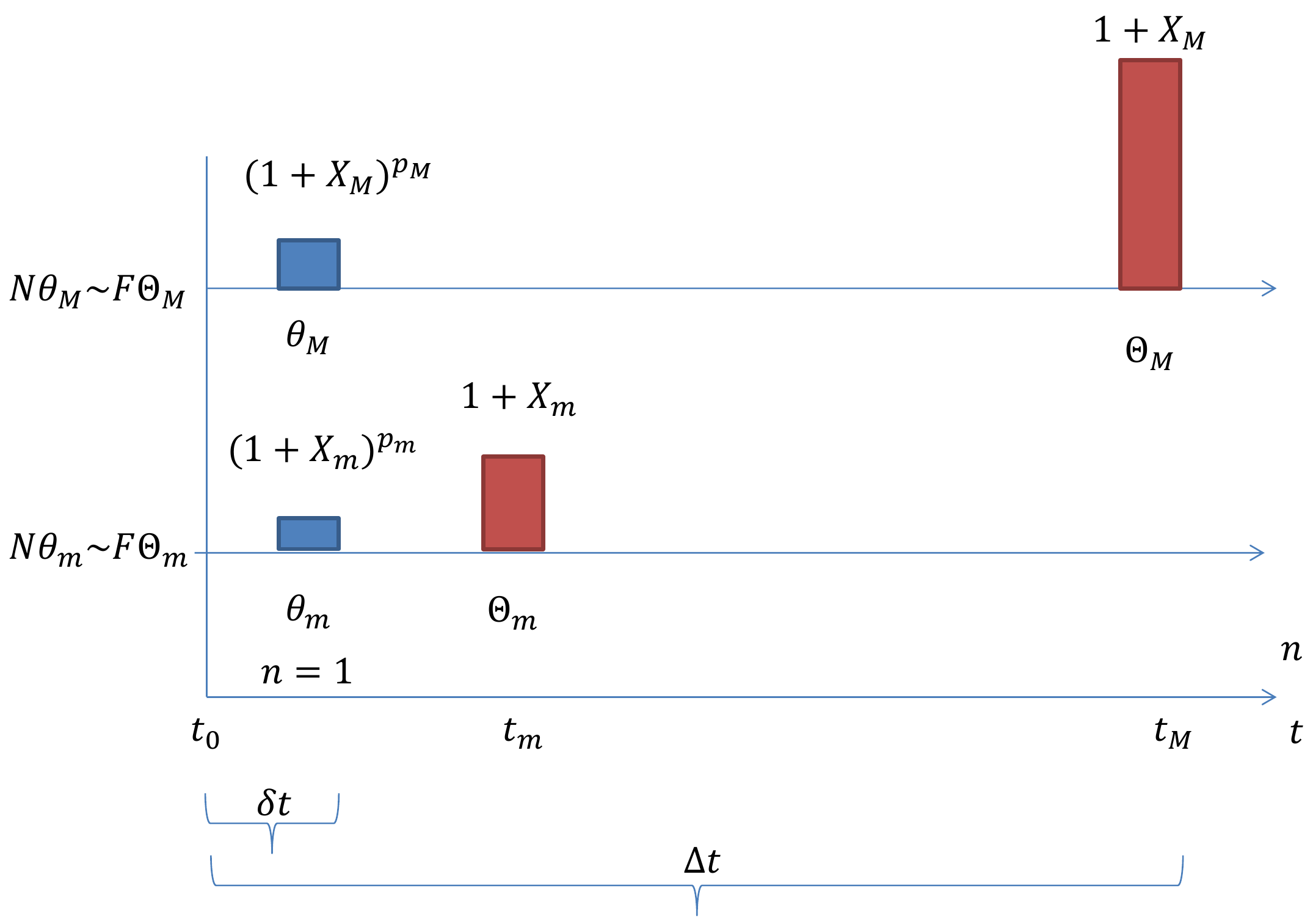}
		\caption{  Intuition about semantics of similarities $N\theta_m \sim F\Theta_m$ and  $N\theta_M \sim F\Theta_M$. }	
	\label{Fig:Semantic}
\end{figure} 

Figure \ref{Fig:Semantic} shows an intuition about the similarities  $N\theta_m \sim F\Theta_m$ and  $N\theta_M \sim F\Theta_M$. 
First, there are two timelines to represent two possible worlds before making a decision. Second, there is no determinism about payments $M$ and $m$.  The hypotheses $\theta_m$ and $\theta_M$ have as result the time averages $(1+X_m)^{p_m}$ and $(1+X_M)^{p_M}$ at the end of the first period, $n=1$. Thus, only $(1+X_m)$ and $(1+X_M)$ represent payments in the future (see  red bars).
Third, although payment of $m$ is  scheduled for $t_0+\delta t$,  the uncertainty can  delay it (if $p_m\leq 1$, then it is possible $t_m \geq t_0+\delta t$). The delay will never occur if $p_m=1$. Fourth, the same reasoning can be applied to the hypothesis $\Theta_M$, but no additional delay is illustrated. This is just to remember that payments can happen at the agreed time. Finally, it should be noted that a lower probability increases the waiting expectation ($p_M<p_m$ makes $t_M>t_m$ more likely).

\section{Decision criterion} \label{DC}

 Note  in  Figure \ref{Fig:Semantic} that the decision can be taken by choosing the largest time average at $n = 1$. Consider  $M$  large enough to 
 \begin{eqnarray}
(1+X_M)^{p_M} &>& (1+X_m)^{p_m} .
\end{eqnarray}
The above inequality is due to the assumption that the hypotheses $\Theta_M$ and $\Theta_m$ have different dynamics.

Now, substitute  the expressions \ref{eq_impactM} and \ref{eq_impactm} to obtain
\begin{eqnarray}
\left(1+\frac{M}{W_0} \right)^{p_M} &>& \left(1+\frac{m}{W_0} \right)^{p_m}.
\label{BhomBawerk}
\end{eqnarray}
The result of the decision criterion is the maximum time average in the next period
 \begin{eqnarray}
D_c  &=& \max \left\{\left(1+\frac{M}{W_0} \right)^{p_M},    \left(1+\frac{m}{W_0} \right)^{p_m}\right\}  \\
&=&  \left(1+\frac{M}{W_0} \right)^{p_M}.
 \label{Decision_Criterion_early}
\end{eqnarray}
In this case, the opportunity to receive $m$ is foregone to receive $M$ later.

\subsection{Preference reversal}

  Preference reversal can be observed when two time-separated options are delayed. Initially, when individuals are asked to choose between one apple today and two apples tomorrow, then  they prefer only one apple today. However, when the same options are long delayed, for example, choosing between one apple in one year and two apples in one year plus one day, then to add one day to receive two apples becomes acceptable \cite{thaler1981some}. Similar  behaviors can be observed in humans \cite{kirby1995preference,green1994temporal,millar1984self,solnick1980experimental} and pigeons \cite{ainslie1981preference,green1981preference}.

The preference reverse problem  helps us to understand the effects that probabilities and amounts have on time averages. So, let's start explaining each effect separately.

First, when two risky payment hypotheses have the same chance, then only the amounts are relevant to the decision.  For example, consider evaluating the hypotheses  $\Theta_M$ = ``I receive $M$'' and $\Theta_{2M}$ = ``I receive $2M$'', where the future tenses  ``It will be the case that I receive $M$ today'' and ``It will be the case that I receive $2M$ today'' are certain events. If the decision maker has to choose between the two hypotheses, then  receiving $2M$ today is preferable because
\begin{equation}
\left( 1+\frac{M}{W_0}\right) < \left( 1+\frac{2M}{W_0}\right).
\label{today1}
\end{equation}
In this case, both hypotheses have the same frequency.  Therefore, when two hypotheses have the same payment probability, $p$, the judgment is also made by choosing the highest payment, 

\begin{equation}
\left( 1+\frac{M}{W_0}\right)^p < \left( 1+\frac{2M}{W_0}\right)^p, \textnormal{ for } 0\leq p \leq 1.
\label{today2}
\end{equation}

 Second,  the delay of payments decreases the frequency and, consequently, decreases the time average.  Now, consider delaying $2M$ until tomorrow. Thus, the decision maker says: ``It will be the case that I receive $2M$ tomorrow''. When the waiting time doubles to receive $2M$,  then the frequency this payment is halved. Thus, receiving $M$ today is preferable because
  \begin{equation}
\left( 1+\frac{M}{W_0}\right) > \left( 1+\frac{2M}{W_0}\right)^{\frac{1}{2}}.
\label{Twice}
\end{equation}
Therefore, delaying a payment  reduces the time average of a hypothesis. 
 
 \begin{figure}[!ht]
	\centering
	\includegraphics[scale=0.65]{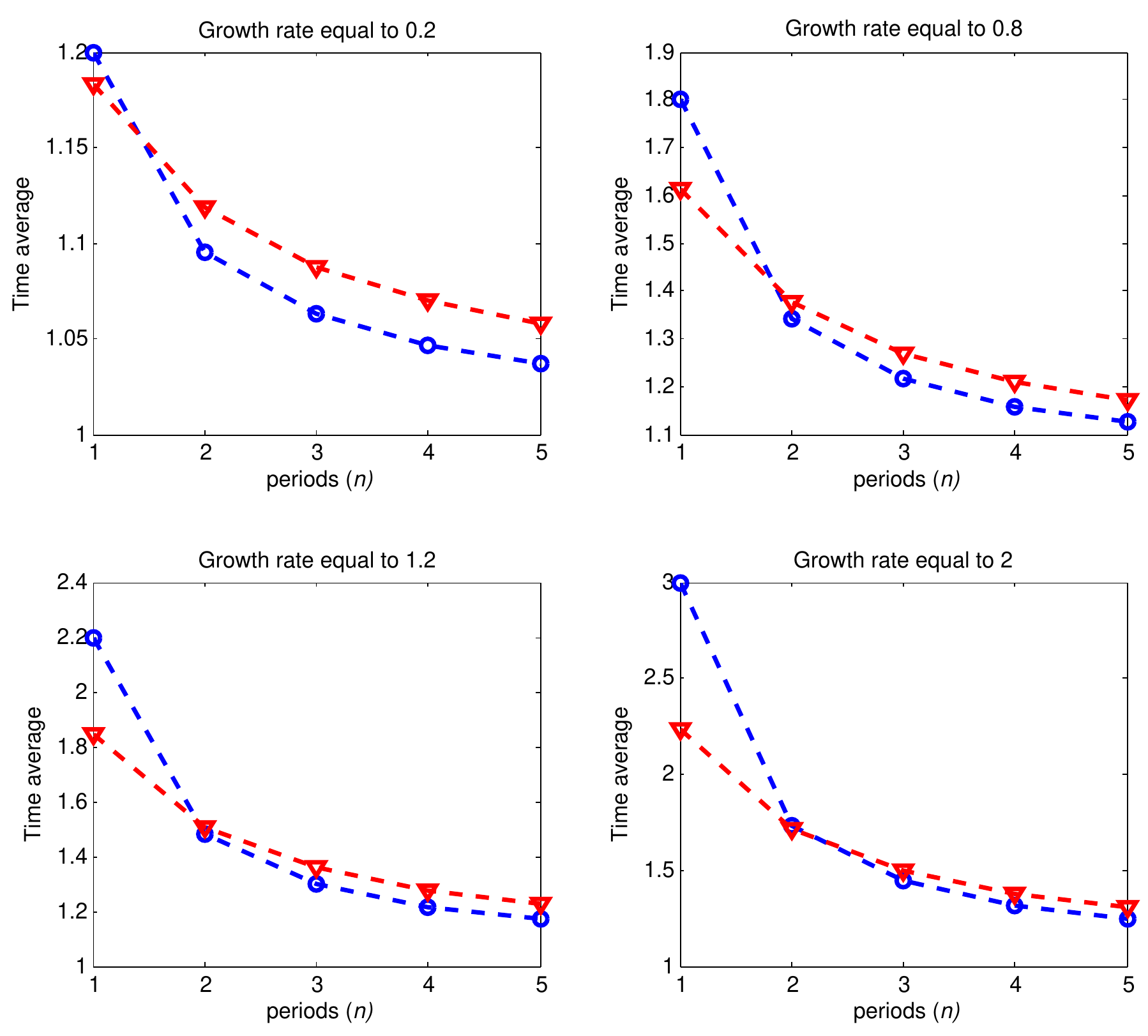}
		\caption{ Preference reversal for different rates $M/W_0$ (0.2, 0.8, 1.2 and 2). The $\circ$-blue curve is the time average for $\Theta_M$ and the $\triangledown$-red curve is the time average for $\Theta_{2M}$. Initially, $\Theta_M$ is preferable because its time average is bigger. However, $\Theta_{2M}$ is preferable after the second period.}	
	\label{figPreferenceReversal}
\end{figure} 
 
 There are cases where amounts and probabilities are simultaneously relevant to the time average.  An example is considering the future tenses  ``It will be case that I receive $M$ in $n$ days'' and  ``It will be case that I receive $2M$ in $n+1$ days''.
 Since the waiting time to receive $ M $ is shorter, then we can realize that the number of trials to receive $M$ will be greater in the future ($(n+1)/n$ trials to receive $M$ for each trial to receive $2M$). Thus, the rational choice between  $ \Theta_M$ and $ \Theta_{2M}$ depends  on choosing the highest  time average: 
\[ D_c=\max \left\{\left( 1+\frac{M}{W_0}\right)^{\frac{1}{n} },\left( 1+\frac{2M}{W_0}\right)^{\frac{1}{n+1}}\right\}.\]
  When $n=1$,  receiving the payment of $M$ is preferable (see equation  \ref{Twice}). However, when $n$ is large enough, the relation $(n+1)/n$ tends to  1 and the payments of  $M$ and $2M$ have almost the same frequency. Now remember the equation \ref{today2}, when two hypotheses have the same probability, the decision criterion chooses the highest payment. Thus, the performances between the hypotheses reverse when the payments are shifted to a longer period. 
  
  Figure \ref{figPreferenceReversal} shows as the  time averages for $\Theta_M$ and $\Theta_{2M}$ are reversed for different rates, $M/W_0$. 
Note that  the payment probabilities are relevant to choosing the lowest amount in the first period, but they lose the relevance in the second period and  the decision is reverted.

\section{A cause of indifference between early and late payments} \label{TCOIBEALP}

The discount problem can be proposed to an individual by asking the following question: What is the $ M $ amount to receive later that makes you indifferent to receiving $ m $ earlier? Therefore, the indifference between early and late payments is the central issue in this problem.

 Initially, for simplicity, consider a uniform probability distribution with no delay. If there is no delay, then remember that $p_m = 1$ and $p_M = 1/n$, where $1/n$ is the probability to receive $ M $ in each period that maximizes entropy (see uniform distribution in subsection \ref{sub:payment_probabilities} for $\alpha = 1$). Thus, the payment of $M$ has a frequency of $1/n$ relative to the payment of $m$, that is
\begin{equation}\frac{p_M}{p_m}=p_M=\frac{1}{n}  \end{equation}

Substituting the probabilities in the decision criterion, we have
 \begin{eqnarray}
\nonumber D_c  &=& \max \left\{\left(1+X_m \right)^{p_m},\left(1+X_M \right)^{p_M} \right\}  \\
&=&  \max \left\{\left(1+\frac{m}{W_0} \right),\left(1+\frac{M}{W_0} \right)^{\frac{1}{n}} \right\}.
 \label{Decision_Criterion_early1}
\end{eqnarray}

Since $M$ is greater than $m$, then we can write $r = M/m$ to obtain:
 
 \begin{eqnarray}
D_c  &=&  \max \left\{\left(1+\frac{m}{W_0} \right),\left(1+r\frac{m}{W_0} \right)^{\frac{1}{n}} \right\}.
 \label{Decision_Criterion_early2}
\end{eqnarray}

\begin{figure}[!ht]
	\centering
	\includegraphics[scale=0.8]{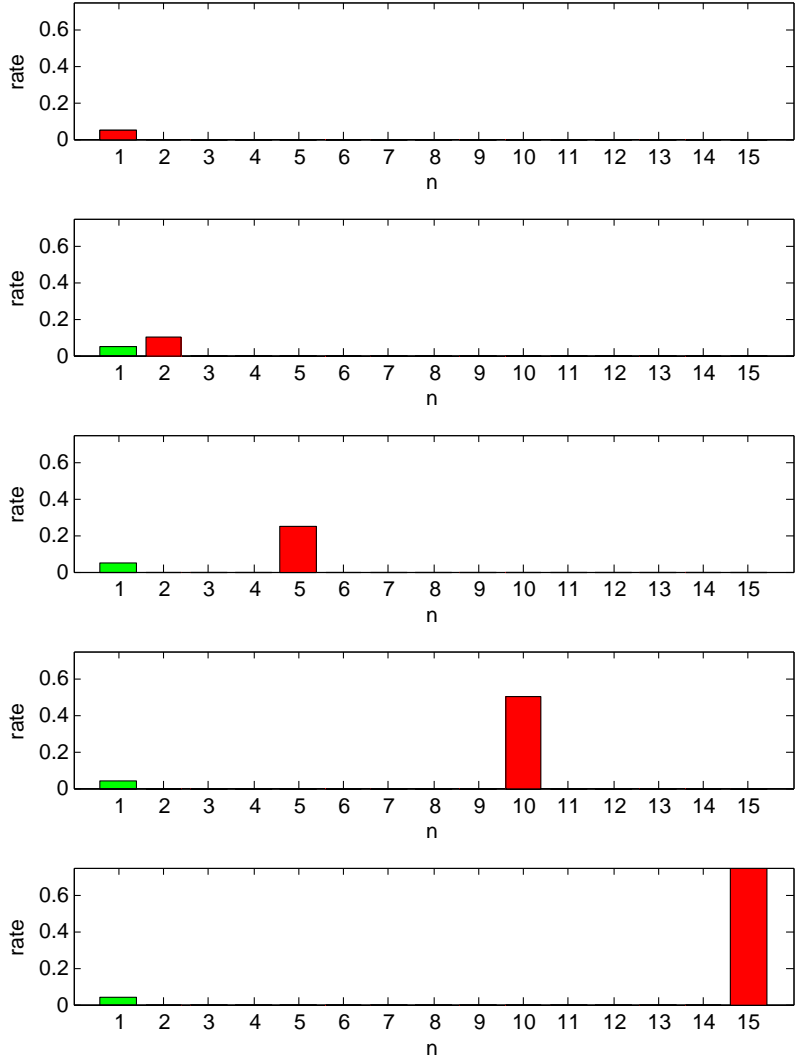}
		\caption{Rates of impact factors and time averages versus $n$. Consider possible payments where there are red bars ($n$ equal to 1, 2, 5, 10 and 15). In the top bar plot, the possible payment of $m$ in the first period is equivalent to 5\% of personal wealth, $X_m=0.05$. In the other bar plot, the red bars are the growth rates $X_M=nX_m$ and the green bars are time average rates in the first period, $(1+nX_m)^{\frac{1}{n}}-1$. Note that the time average rates are almost equal to $X_m$ at $n=1$. }	
	\label{ManyWorlds}
\end{figure}

Now, the individual must be indifferent to receive $m$ early or $M$ later. For both to be equal at $n = 1$, we can propose  $r = n$, 
 \begin{eqnarray}
D_c  &=&  \max \left\{\left(1+\frac{m}{W_0} \right),\left(1+n\frac{m}{W_0} \right)^{\frac{1}{n}} \right\}.
 \label{eq:indiferent}
\end{eqnarray} 
If the payment amount is much less than personal wealth, $m\ll W_0$, then these time averages remain close to other values of $n$ (see Figure \ref{ManyWorlds}).  Particularly, if $0<X_m< 1$ and $n\geq 1$, then 
\[1\leq \left(1+nX_m \right)^{\frac{1}{n}} \leq 1+X_M.\]
Now, if $0<X_M\ll 1$, then  we have the following causal relationship 
\begin{equation} 0<X_m\ll 1 \quad\mbox{and}\quad   \frac{1}{n}=\frac{p_M}{p_m}=\frac{X_m}{X_M} \Rightarrow \frac{\left(1+nX_m \right)^{\frac{1}{n}}}{1+X_m}\approx 1, \;\; \forall \;\; n\geq 1.
\label{Expresion:GrowthStraight}
\end{equation}
 
Finally, calculating the time average after $n$ periods to obtain $1+X_M$, then the later payments grows linearly with the number of periods, that is, 
\[1+X_M=\left[(1+nX_m)^{\frac{1}{n}}\right]^n \Rightarrow M=nm,\;\; \forall \;\; n\geq 1.\]

Figure \ref{ManyWorlds} shows possible payments in  $n=1$, 2, 5, 10 and 15 periods. At $n=1$, the payment has a growth rate $X_m=0.05$ (red bar on the top bar plot). In the other bar plots, the rates of impact factors increase linearly $X_M=nX_m$ (red bars). However, we visually perceive that the rates of time average are almost equal to $X_m$ (compare green bars with the red bar at $n=1$). This illustrates how the linear growth of these payments generates time averages very close.

\subsection{The initial payment problem}

Expression \ref{Expresion:GrowthStraight} considers $M=0$ for  $0 \leq n <1$.  To solve this problem, we can assume there has already been a payment $m_0$ at $t_0$. Thus, we can write
\begin{equation}\frac{m_0+W_0}{M+W_0}=\frac{1}{1+\rho n},\label{DescontoHiperbolicoTradicional}\end{equation}
where  $M=m_0$  at $t_0$ ($n=0$). This  phenomenon has been documented  as hyperbolic discounting \cite{herrnstein1981self, mazur1987adjusting}. 

For the calculation of a per-period discount rate we can find
\begin{equation}
    \rho=\frac{m-m_0}{W_0+m_0},
    \label{eq:discountrate}
\end{equation}
where $m$ is the payment in first period ($n=1$). Consequently, we have 
\begin{equation} M=m_0+ (m-m_0)n.\label{eq:M}\end{equation}

Note that although the amount shows growth independent of personal wealth (equation \ref{eq:M}), the discount rate depends on it (equation \ref{eq:discountrate}).

\subsection{Relative frequency between early and late payments with possibility of delays}
  
In section \ref{SLT}, the similarities $N\theta_m \sim F\Theta_m$ and $N\theta_M \sim F\Theta_M$ were found. Now, the objective is to find the similarity \[F\Theta_m \sim F\Theta_M.\]
In other words, if $m$ and $M$ are early and late non-deterministic payments in the future, we need a rule that makes them similar. 

If the payment of $M$ can be delayed beyond the deadlines, then the $P_M$ function increases its entropy reaching the value 1 later (see uniform distribution in subsection \ref{sub:payment_probabilities} for $\alpha >1$). The same rule can be applied to $P_m$. As these delays can be disproportionate, then a factor $q-1$ is needed to adjust the relative frequency, 
\begin{equation}\frac{p_M}{p_m}=\frac{1}{n(q-1)}  \;\; \forall \;\; q>1,\,n\geq 1.  \end{equation}

By analogy to expression \ref{Expresion:GrowthStraight}, if we assume the following cause of similarity
\begin{equation} 0<X_m\ll 1 \quad\mbox{and}\quad   \frac{1}{n(q-1)}=\frac{p_M}{p_m}=\frac{X_m}{X_M},
\label{Expresion:GrowthStraight2}
\end{equation}
then the decision criterion result in 
\begin{eqnarray}
\nonumber    D_c &=&   \max\left\{ (1+X_m)^{p_m},   (1+X_M)^{p_M}\right\}\\
   &=& \max\left\{ (1+X_m)^{p_m},   [1+(q-1)nX_m]^{\frac{^{p_m}}{(q-1)n}} \right\}.
     \label{eq:indiferentHipebolica}
\end{eqnarray}
Particularly, if  $p_m=1$ and $q=2$, then we have the expression  \ref{Expresion:GrowthStraight} again. 

For there to be an indifference between early and late payments in the expression \ref{eq:indiferentHipebolica}, then time averages must be equal. However, this criterion is very strict for discounters. In practice, individuals have low accuracy to compare approximate averages. Therefore, it is only necessary that the averages have low contrast for all $n\geq 1$.

\subsection{Contrast between time averages}
\label{CBTA}

We can make an analogy with information processing, where the processing efficiency depends on the contrast between the powers of signals. Therefore, if the time averages act as signals to stimulate a response from the discounter, then the contrast between their amplitudes is given by:
\begin{equation}
    CR_{dB}=-20\,p_m\; \log_{10}\frac{[1+(q-1)nX_m]^{\frac{1}{(q-1)n}}}{1+X_m}.
    \label{CR}
\end{equation}

\begin{figure}[!ht]
	\centering
	\includegraphics[scale=0.8]{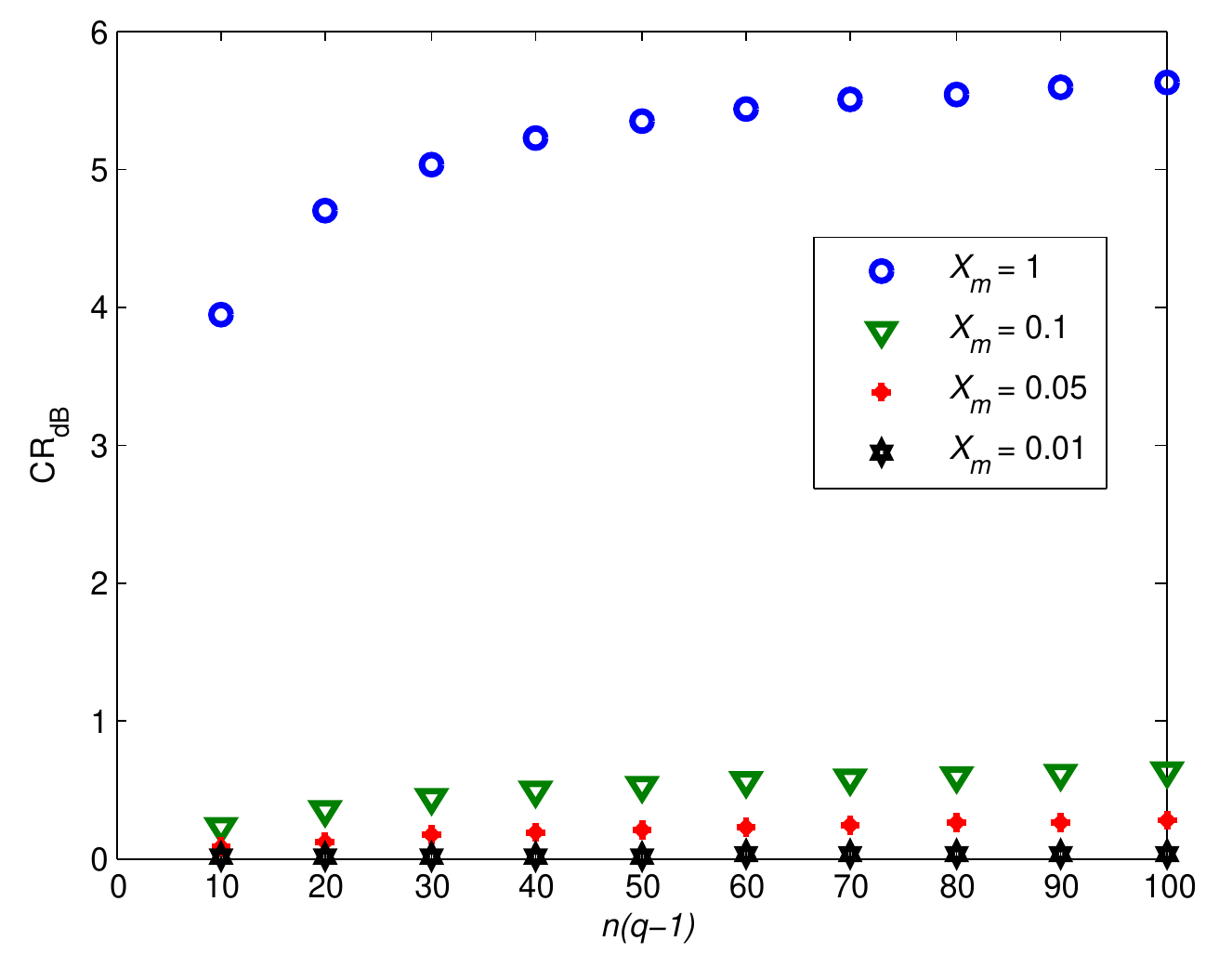}
		\caption{Contrast ratio between time averages  according to equation \ref{CR} ($p_m=1$ and $X_m$ = 0.01, 0.05, 0.1 and  1). }	
	\label{contrast}
\end{figure}

Figure \ref{contrast} shows the contrast ratio between time averages. For payments  less than  10\% of personal wealth, the contrast ratio does not exceed $ 0.65\, dB$ during $100/(q-1)$ periods. This is true even for exponential discounters, or rather when $q\to 1^+$ and $0<X_m\ll 1$, then we have $1+X_m\approx e^{X_m}$ (Taylor approximation). 
However, when the payment is equivalent to 100\% of personal wealth, then the contrast ratio increases significantly.  It reaches almost $4 \, dB$ in the first $10/(q-1)$  periods and exceeds the $5\, dB$ after $30/(q-1)$  periods. Under these conditions, the indifference between early and late payments can be lost, leading the individual to look for another discount method. Therefore, the rule of indifference in the expression \ref{Expresion:GrowthStraight2} is valid only for payments much smaller than the individual's wealth.

\subsection{Hyperbolic discounting}

Finally, the time average  has the following form after $n$ periods
\begin{eqnarray}
\nonumber (1+X_M)^{np_M}=1+\frac{\tilde{M}}{W_0} &=&\left\{[1+(q-1)X_m n]^{\frac{p_m}{(q-1)n}} \right\}^n \\
&=&[1+(q-1)X_m n]^{\frac{p_m}{(q-1)}},\; \forall \; n,q \geq 1,
\label{q_exponentialDiscounting}
\end{eqnarray}
where $\tilde{M}$ is the  expected amount after $n$ periods when $m$ and $M$ can be delayed. Now, $\tilde{M}$ does not always grow linearly with $n$.

Similarly the equation \ref{eq:discountrate}, a discount rate $\rho$ can be assumed in place of $X_m$ to describe payments from  $t_0$ ($n=0$). Therefore, the discounting of time average after a delay of $n$ periods can be calculated by
\begin{eqnarray}
\nonumber \frac{W_0+m_0}{\tilde{M}+W_0}&=&[1-(1-q)\rho n]^{\frac{p_m}{(1-q)}} \\
&=& [e_q^{-\rho n}]^{p_m} \; \textnormal{  for } q \geq 1,\; n\geq 0,
\label{q_exponentialDiscounting2}
\end{eqnarray}
where $\tilde{M}=m_0$ when $n=0$.

When payments are very likely in the first period, one can consider $p_m\approx 1$. In these cases, the discount function above has fitted the time preference of discounters \cite{benhabib2010present,benhabib2004hyperbolic, tanaka2010risk}.

\section{Simulation of a hyperbolic discounter}
\label{SOADM}

\begin{figure}[!ht]
	\centering
	\includegraphics[scale=0.8]{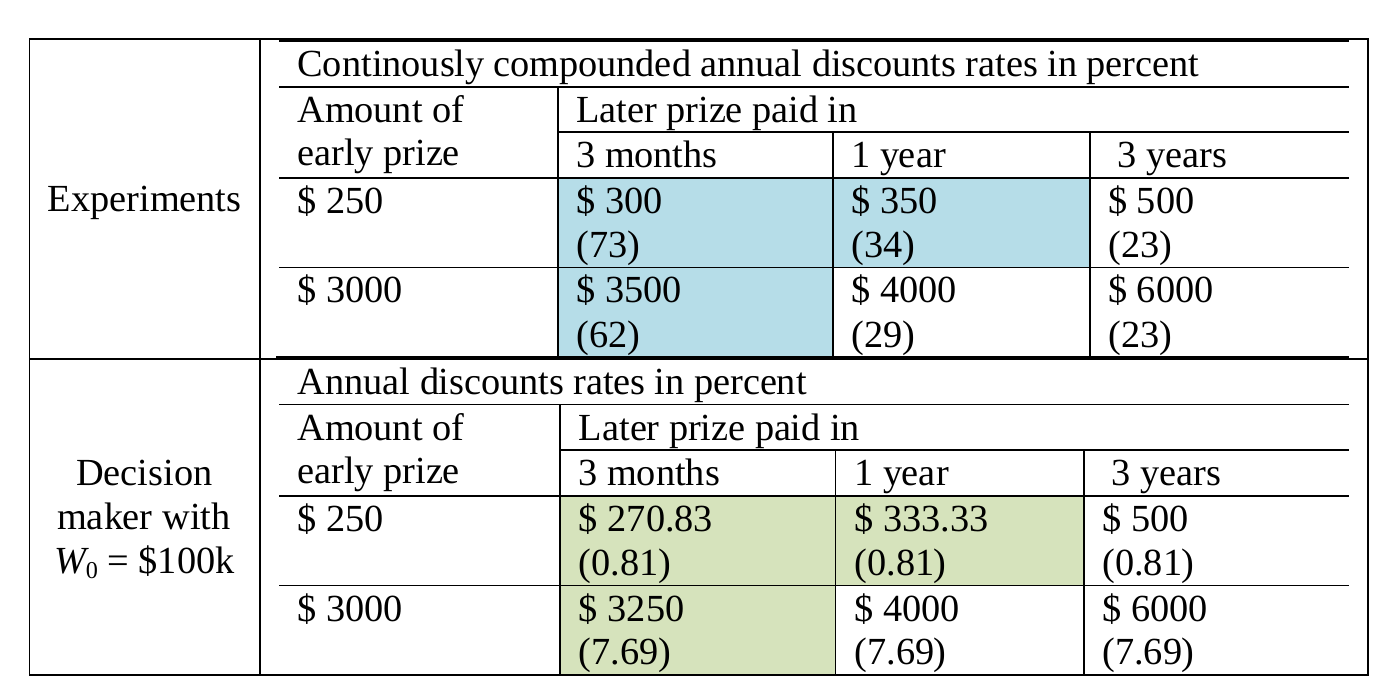}
		\caption{A simulation of a discounter versus the discounters of Thaler's experiments. First part of table, there are results of two Thaler's experiments, where the median of responses were reported \cite{thaler1981some}.  The second part of Table, a discounter is simulated with  personal wealth $W_0 = \$ \; 100k$ and $q=2$, where he participates in both experiments. }	
	\label{figDiscounting4}
\end{figure}

 Now, we need to understand the discount rates according to equation \ref{eq:discountrate}. For this, a discounter is simulated in two Thaler's experiments \cite{thaler1981some} (see Figure \ref{figDiscounting4}). These experiments did consist of communicating to the participants that they had won a prize in a lottery held by their bank (\$ 250 for experiment 1 and \$ 3000 for experiment 2).  So that, they were asked how much they would require to make waiting just as attractive as getting the money now. Then, the responses were given and their median  were reported (see Figure \ref{figDiscounting4}).  

Initially, we should note that the amounts of simulated discounter are almost the same observed in Thaler's experiment. For this, we can compare where the amounts are different (light blue and green regions in Figure \ref{figDiscounting4}). Only discount rates show significant differences, therefore,  some explanations are necessary about them.

\subsection{Constant discount rates}

 In Thaler's experiment, the rate  decreases over time. The reason is  the annual discount rate compounded continuously. When the {\it hyperbolic discount rates}\footnote{The equation \ref{eq:NewFormaRates} is deduced from the hyperbolic discounting (equation \ref{q_exponentialDiscounting2}). For $q = 2$ and $p_m=1$, the equation \ref{eq:NewFormaRates} has the form of the equation \ref{DescontoHiperbolicoTradicional}, where the per-period discount rate  is given by the equation \ref{eq:discountrate}.} are calculated through 
\begin{eqnarray}
\nonumber \rho &=&-\frac{1}{n}\frac{\left[\left(\frac{W_0+m_0}{W_0+\tilde{M}}\right)^{\frac{1}{p_m}}\right]^{1-q}-1}{1-q}\\
&\equiv  & -\frac{1}{n}\ln_q\left(\frac{W_0+m_0}{W_0+\tilde{M}}\right)^{\frac{1}{p_m}}
\label{eq:NewFormaRates}
\end{eqnarray}
for $q=2$ and $p_m=1$, then the rates remain constant for the same discounter after 3 months, 1 years and 3 years (0.81\% for experiment 1 and 7.69\% for experiment 2). 

\subsection{Reversing the magnitude effect}

In Thaler's experiment, lower discount rates are found for larger amounts. This phenomenon is called magnitude effect. However,  this effect is reversed when we use the equation \ref{eq:NewFormaRates} to calculate the discount rates. Now, they are growing along with the amounts $m_0$. More specifically, the same discounter with a personal wealth of \$ 100k is simulated in both experiments, but he has an annual discount rate of 0.81\% when discounting \$ 250 and 7.69\% when discounting \$ 3000.  Therefore, increasing the amount to be discounted has increased the discount rate.

\section{The influence of personal wealth on hyperbolic discounting}
\label{DATWOTDM}

 In this paper, the indifference between early and late payments in hyperbolic discounting is caused by payments much less than the personal wealth. As we can see in Figure \ref{figDiscounting4}, the discount rates of the simulated discounter are much lower than those calculated by Thaler (0.81\% less than 73\%, 34\%, 23\% and  7.69\% less than 62\%, 29\%, 23\%). The reason is the personal wealth in the simulation (100 thousand dollars). Thus, the amounts are limited to 6\% of wealth to preserve the indifference between early and late payments. 

Evidence of personal wealth interference in hyperbolic discount experiments can be found in the literature. For example, if $\rho$ depends on  $W_0$ (see equation \ref{eq:NewFormaRates}), then there must be great variability among discount rates of individuals. The reason is  power-law probability distribution of wealth, where $W_0$  can vary abruptly from one individual to another \cite{levy1997new,druagulescu2001exponential,sinha2006evidence,klass2006forbes}.    For example,  a ``wide variation among persons'' was mentioned by Thaler \cite{thaler1981some}. Moreover, in \cite{frederick2002time} was presented a tremendous variability in the estimates of literature.  Later,  the experiments focused on investigating individual discounters because it was more productive to recognize heterogeneity among the individuals \cite{benhabib2004hyperbolic, benhabib2010present,akin2007experimental}.  
 
 In addition, the discount rate is influenced by individual states of scarcity and abundance of goods.  According to equation \ref{eq:NewFormaRates}, if a discounter is poor (small $W_0$), then he places a higher preference (great $\rho$). On the other hand, if he is rich (great $W_0$), then he has a small $\rho$. An experiment that supports this deduction is in  \cite{tanaka2010risk}, where groups were evaluated by income. Then, it was found that mean income is correlated with the discount rates. More specifically,  people living in wealthy places have a lower hyperbolic discount rate. Therefore, the individual states of scarcity and abundance of goods has a strong influence on time preference.    
 
 Finally, when personal wealth is much greater than the first payment future ($0<X_m\ll 1$), then time average it does not change significantly with the delays. Thus, if individuals do not perceive the change, then they may allow several payment dynamics. For example, although time averages assume a multiplicative dynamic (optimal wealth growth),  payments with low impact on wealth allow an additive dynamics\footnote{Make $q=2$ and $p_m=1$ in the decision criterion (see expression  \ref{eq:indiferentHipebolica}).}  because the time average does not change significantly after $n$ periods
\begin{equation}
\underbrace{(1+X_m)}_{\text{First period}}   \approx \underbrace{\left(1+nX_m \right)^{\frac{1}{n}}}_{n\text{-th period}}.
 \label{eq_linear_average}
\end{equation} 
  Similarly, a {exponential discounter}\footnote{Make $q\to 1$ in the decision criterion (see expression  \ref{eq:indiferentHipebolica}).} also has time averages very close, 
\begin{equation}
\underbrace{(1+X_m)^{p_m}}_{\text{First period}}   \approx \underbrace{e^{p_mX_m}}_{n\text{-th period}}.
\label{eq_linear_exponential}
\end{equation} 
 Under theses conditions, the  $q$-exponentials interpolate future payments between hyperbolic and exponential functions without relevant losses for the individual.  Therefore,  a  cause for the hyperbolic discounting    is  the indistinguishability between time averages when the payments have low impact on personal wealth. This means that the causes of payment dynamics may be more physical than psychobehavioral.

\section{Conclusion}
\label{Conclusion}

This paper considers the calculation of time averages to solve intertemporal choice problems, where maximum uncertainty,  impact on personal wealth and accuracy are relevant to making a decision.

A statistical interpretation of the preference reversal problem can be obtained through that approach. More specifically, if a large amount is always overdue 1 period of the small amount and both are delayed over time, then
\begin{enumerate}
    \item payment probabilities and amounts are relevant to making the decision when the delay is short;
    \item but when the delay is long, the payment probabilities are almost the same, so only the amounts to be paid become relevant to make the decision.
\end{enumerate}
Therefore, if the payment probabilities are relevant to choosing the lowest amount in the short term, they lose that relevance in the long term and the decision is reverted.

Also, non-exponential discounting can be applied  when the payments are irrelevant to personal wealth. The discount problem arises when the individual asks himself: How much should I receive later to be indifferent to receiving $m$ earlier? Note that the indifferent needs a rule that guarantees a similarity between early and late payments. In this context, the following rule is valid when human beings  do not accurately distinguish small averages:
\begin{enumerate}
    \item if the amounts are much less than the personal wealth, then the $q$-exponential discounting guarantees the indifference between early and late amounts in several future periods.
\end{enumerate}
Consequently, there is a rational discounter that allows non-exponential discounting in the face of these conditions.

Finally, although time averages assume a multiplicative dynamics (optimal wealth growth),  payments with low impact on wealth allow an additive dynamic ($q=2$). Thus, the  $q$-exponentials interpolate future payments between hyperbolic and exponential functions because small payments make the time averages indistinguishable over time. This means that   there are physical causes that allow explaining the hyperbolic discounting regardless of psycho-behavioral assumptions.

\section{Acknowledgements}
First, I would like to thank the reviewers for their relevant comments that
improved this manuscript. Second, I declare this  study  was  financed  in  part  by  the  CAPES - Brasil - Finance  Code  001.

\bibliographystyle{unsrt}
\bibliography{sample}

\newpage
\begin{table}[h!]
\centering
\caption{List of Symbols}
 \begin{tabular}{l l} 
 \hline
 Symbol & Name and interpretation  \\ [0.5ex] 
 \hline
 $M$ & amount to be received after a long period \\
 $m$ & amount to be received after a short period \\
 $t_m$ & future time to receive $m$ \\ 
 $t_M$ & future time to receive $M$  \\ 
 $t$ & any future time  \\ 
 $t_0$ & present time  \\
 $m_0$ & amount  received at the instant $t_0$ \\
 $\delta t$ & short waiting period to receive $m$  \\
 $\Delta t$ & long waiting period to receive $M$\\
 $n$ & number of periods $\delta t$ until the scheduled time to receive $M$  \\
 $\Theta_m$ & hypothesis: ``I receive $m$'' \\
 $\Theta_M$ & hypothesis: ``I receive $M$'' \\
 $\Theta_{2M}$ & hypothesis: ``I receive $2M$'' \\
 $\theta_m$ & hypothesis: ``I receive an amount less than $m$'' \\
 $\theta_M$ & hypothesis: ``I receive an amount less than  $M$'' \\
 $W_0$ &  personal wealth of decision maker  at instant $t_0$  \\
 $X_M$  &  growth rate associated with the hypothesis $\Theta_M$, $M/W_0$\\
 $X_m$  &  growth rate associated with the hypothesis $\Theta_m$, $m/W_0$\\
 $x_M $  & growth rate associated with the hypothesis $\theta_M$ \\
 $x_m$  &  growth rate associated with the hypothesis $\theta_m$ \\
 $F$  &  modal operator that means ``It will at some time be the case that...''\\
 $G$  &  modal operator that means ``It always will be the case that...''\\
 $N$  &  next operator of temporal logic \\
 $u$  &  total time that $\theta_M$ is true according to $G\theta_M$ \\
 $\tau(u)$ & total time that $\Theta_M$ is true according to $GF\Theta_M$\\
 $p_M$  &  payment probability of $M$ at the end of the first period, $t_0 + \delta t$ \\
 $p_m$  &   payment probability of $m$ at the end of the first period, $t_0 + \delta t$ \\
 $D_c$ & maximum time average from the decision criterion\\
 $q$  &   adjustment between the relative frequencies when $m$ and $M$ can be delayed\\
 $\rho$  &  per-period discount rate\\ 
 $\tilde{M}$ & expected amount after $n$ periods when $m$ and $M$ can be delayed\\[1ex] 
\end{tabular}
\label{table:1}
\end{table}






% \end{thebibliography}

\end{document}